# Fractional Order Modeling of a PHWR Under Step-Back Condition and Control of Its Global Power with a Robust PI$^\lambda$D$^\mu$ Controller

Saptarshi Das, Shantanu Das, and Amitava Gupta

*Abstract*—Bulk reduction of reactor power within a small finite time interval under abnormal conditions is referred to as *step-back*. In this paper, a 500MWe Canadian Deuterium Uranium (CANDU) type Pressurized Heavy Water Reactor (PHWR) is modeled using few variants of Least Square Estimator (LSE) from practical test data under a control rod drop scenario in order to design a control system to achieve a dead-beat response during a stepped reduction of its global power. A new fractional order (FO) model reduction technique is attempted which increases the parametric robustness of the control loop due to lesser modeling error and ensures iso-damped closed loop response with a PI$^\lambda$D$^\mu$ or FOPID controller. Such a controller can, therefore, be used to achieve active step-back under varying load conditions for which the system dynamics change significantly. For closed loop active control of the reduced FO reactor models, the PI$^\lambda$D$^\mu$ controller is shown to perform better than the classical integer order PID controllers and present operating Reactor Regulating System (RRS) due to its robustness against shift in system parameters.

## I. INTRODUCTION

RAPID reduction of the bulk power in a nuclear reactor is generally done under load following operations [1] or under some abnormal operating conditions. Control rods are inserted to a pre-specified level for rapid reduction of the global power in operating Indian PHWRs which is known as *step-back* and simultaneously the set point of the demand power is gradually reduced. The step-back action creates a power undershoot while producing a sluggish response as can be found with other means of sudden negative reactivity insertion in the PHWR [2]-[5]. The safety constraints do not permit excessive power oscillation in the nuclear reactor [6], since at very low power undershoot, the reactor may get poisoned out due to sudden reduction in the thermal neutron flux caused by control rod drop [7].

This work has been supported by Department of Science & Technology (DST), Govt. of India under PURSE programme.
S. Das and A. Gupta are with School of Nuclear Studies & Applications and also with Department of Power Engineering, Jadavpur University, Salt Lake Campus, Kolkata-700098, India (E-mail: saptarshi@pe.jusl.ac.in, amitg@pe.jusl.ac.in).
Sh. Das is with Reactor Control Division, Bhabha Atomic Research Centre, Mumbai-400085, India (E-mail: shantanu@magnum.barc.gov.in).

From the point-kinetic governing equation of PHWRs [7], [2]-[4], it is evident that the reactivity (equivalent control rod worth) to power equation is nonlinear in nature. One approach to analyze the dynamics of such reactors is to use linearized models corresponding to different power levels. Though Indian PHWRs are designed to operate as base load stations, they may be operated at reduced power as base load station. As a result, the linearized models of a reactor gets changed depending on the operating conditions i.e. the initial power level at which a step-back is initiated in addition to the level of control rod drop [8], [9]-[10]. Thus, at different levels, step-back would require a different controller.

Attempts have been made in [1], [9]-[10] to design PID controller considering different linearized transfer function models of the same nonlinear reactor around different operating points. Liu *et al.* [1] designed different controllers which can be switched in accordance with change in the operating point to cope with the reactor nonlinearities, though it is infeasible for a wide variation in operating point like initial power, rod position or both. Also, with a set of switched controllers the stability may not be guaranteed if the operating point lies somewhere in between the discrete modes where the design was attempted. Talange *et al.* [3] and Shimjith *et al.* [4] developed spatial models of large PHWRs and linearized the nonlinear point kinetic equations around a specific steady-state operating point to design the controllers. This may result in some undesirable closed loop dynamics with the same controller for some other operating point because the reactor model gets changed to a large extent in cases that have not been considered in the controller design in [2]-[4].

Further, as in [1], different linear transfer function models have been developed from point kinetics considering different initial power of the reactor but such a governing equation based approach does not take care of the practical issues of modeling uncertainties like geometric considerations [2], material properties, presence of noise in the measured data etc [11]. Regression based techniques have been used by many contemporary researchers to develop models of nuclear reactor to produce PID controllers as in [9]-[10]. The present work, therefore, uses system and noise model identification based approach for reactor model development from practical test-data using regression based process modeling techniques. In order to do this, the reactor is first identified using the dynamics of power variation during a step-back with the change in control rod position as the input and the global reactor power as the controlled variable. It is a well known fact that to achieve a fairly accurate model regression based techniques result in higher order transfer functions which render controller design difficult [12]. Reducing these with a fractional order template and designing a fractional order controller to control the same is an established approach [13].

The physical justification for opting FO modeling strategy for the reactor is discussed next. Recent research [14]-[17] shows the neutron diffusion equation in nuclear reactor is non-Fickian in nature and the reactor point kinetic equations can be accurately described using FO differential equations (the fractional order appearing in space as well as time). So, the motivation of the present work is based on the fact that FO description of the reactor dynamics is a more realistic scenario for neutron diffusion equation and point reactor kinetics.

The present work aims at robust control of the PHWR with nonlinear power dependent dynamics for varying amounts of bulk power reduction at different initial power levels using linearized models and their reduction as varying gain systems which can be efficiently handled with a $PI^\lambda D^\mu$ controller due to its iso-damping property. The idea is to search for linear integer order model first that best represent the reactor nonlinearity around an operating point and then with their most accurate reduced parameter model design a robust FOPID controller that ensures dead-beat power tracking at other operating points also. The motivation behind using a FOPID controller therefore arises from two reasons viz. ease of designing controllers for higher order systems resulting from regression based process modeling techniques and for its iso-damping property which enables use of a single controller for wide range of variation in operating point.

Recently, Saha *et al.* [8] proposed an active rod drop with a robust controller comprising of a FO phase shaper and a Linear Quadratic Regulator (LQR) based PID for an active motor driven step-back. The approach, presented in this paper is an improvement over the active step-back in [8] and in this case the actuator can be driven by a single robust $PI^\lambda D^\mu$ controller. Moreover, the approach presented in [8] achieves phase flattening around the gain cross-over frequency with reduction in phase margin. This limitation is removed in the present approach. A drawback of active step-back mechanism is that it lacks the essential safety feature for power failure in the nuclear power plant and this can be overcome by putting shut-off rods in the reactor to ensure safety, as reported in [18]. In this paper, unlike different controller design for different linearized models in [1], a single robust $PI^\lambda D^\mu$ controller is proposed, for an active motor driven step-back. This improvement is capable of producing satisfactory closed loop response for varying operating points, with respect to that designed in [9], [10] due to enhanced parametric robustness offered by the FOPID controller. For this purpose, identification of the nonlinear reactor is first carried out with a suitable estimator using practical input-output data from plant at a particular power level and the resultant models are then reduced in a flexible template (to retain the dominant dynamic behaviors) for controller tuning. The robust FO phase shaper design for active step-back by Saha *et al.* [8] used conventional AutoRegressive eXogenous (ARX) estimator to model a 500MWe PHWR at 30% rod drop with different initial powers and then reduced it to First Order Plus Time Delay (FOPTD) and Second Order Plus Time Delay (SOPTD) templates. In the present study, the system identification is carried out at various levels of initial power with two different rod drop levels (30% and 50%) using other additional variants of LSE viz. AutoRegressive Moving Average eXogenous (ARMAX), Box-Jenkins (BJ), Output-Error (OE) etc., considering different structures for the system and noise model. A brief comparison of the achievable accuracies with these estimators is also enumerated. The model reduction is attempted in this paper with new *flexible order* templates namely Non-Integer Order Plus Time Delay (NIOPTD) of first and second kind. Reduced order modeling with these NIOPTD templates enhances the robustness of a $PI^\lambda D^\mu$ controller compared to those designed with the classical FOPTD and SOPTD templates. These FO templates allow the order of the reactor model to take any arbitrary non-integer value to capture the delicate transient behaviors with higher accuracy while also extracting the inherent time delays in the identified models. A detailed treatment of the NIOPTD template based model reduction and PID/FOPID controller tuning has been presented in [13]. Reducing higher integer order models (with non-linearity) as compact FO models and their use in FO controller design is an emerging research domain [13], [19] and has been applied in this paper to address a specific problem in nuclear domain i.e. handling power undershoot for passive step-back initiated at low powers. The advantage of reduced parameter NIOPTD model based tuning of FOPID controllers over that with SOPTD templates for reactor control in Saha *et al.* [8] is also highlighted.

Iso-damped tuning of $PI^\lambda D^\mu$ controllers, proposed in [13] is applied to obtain a dead-beat power transient under step-back condition using the reduced NIOPTD plant having maximum dc-gain. It is found that the open loop plant comprising the fastest reactor model and the $PI^\lambda D^\mu$ controller produces a wide range of flatness in the phase curve, so that a change in linearized plant transfer function can be easily handled. The PHWR with its power regulator in closed loop with the thermal feedback is taken as the system and is controlled by a master $PI^\lambda D^\mu$ controller with the master controller output acting as the local set point (Fig. 1).

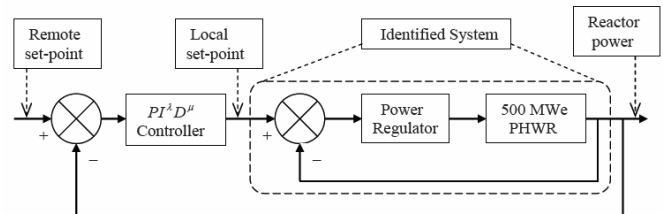

Fig. 1. Identified system comprising of the PHWR along with its power regulator and the proposed modifications in the RRS.

The $PI^\lambda D^\mu$ controller is designed using five independent frequency domain specifications as in [13]. The methodology uses a simultaneous nonlinear equation solving technique to produce a controller which flattens the phase curve for a wide frequency spread around a specified gain cross-over frequency ($\omega_{gc}$) and with phase margin ($\phi_m$) while meeting the user defined specifications on the magnitude of complementary sensitivity and sensitivity functions. The need for iso-damped response in a nuclear reactor is not only to handle the changing transfer function of the nonlinear plant with initial power and rod drop level but also to allow an increase in the gain of the FOPID controller to get a considerably faster control action than the gravity for control rod insertion through the viscous medium, while keeping the power undershoot at same level.

The rest of the paper is organized as follows. Section II discusses about the reactor model development with different variants LSE. The new *flexible order* model reduction templates are introduced in section III for reduction of the identified higher order models of the reactor. Section IV designs a robust PID and a PI$^\lambda$D$^\mu$ controller with the accurate NIOPTD reduced order models of the reactor. Simulated results with the designed FOPID controller are presented in section V for the test reactor considering an active step-back up to 30% and 50%. The paper ends with the conclusion as section VI, followed by the references.

## II. REACTOR MODEL DEVELOPMENT

In a typical 500MWe CANDU type Indian PHWR, the reactor power is controlled by the RRS which consists of mainly three components viz. control rods (CR), adjustor rods (AR) and zone control compartments (ZCC). The ARs are provided for fast startup of the reactor. The CRs are provided for coarse control and ZCCs for fine control of the power level. The set-point of the RRS is specified at the desired level, which is called the demand power and the control loop error is corrected by the continuous measurement of the reactor power level or bulk power by the Self-Powered-Neutron-Detectors (SPND). So, the goal of the reactor control system is to minimize the effective power error (EPE) which is the sum of the difference between demand power and bulk power and the difference between their instantaneous rates [20], [21]. The present RRS in operating PHWRs uses proportional controller as described in [20], [21]. Now, in order to identify the test PHWR from test data along with its power regulator in loop (Fig. 1) as stable transfer function models, the basic philosophy of LSE based system identification techniques and few of its variants along with their relative potentials are first introduced.

### A. System Identification Using Basic LSE

System identification deals with modeling of dynamic systems without prior knowledge of the system's governing physical laws. It is basically finding an approximate model from an input-output experimental data, where the modeling requires less insight of the actual system physics. There are several classical identification methods e.g. time response based, frequency response based methods etc [22], [23]. In the present work, a time response based system identification approach is adopted, to find out the transfer function between power developed by a nuclear reactor and level of control rod drop for a 500MWe PHWR of CANDU type.

If it be assumed that at time event $t$, the input and output of an unknown system are $u(t)$ and $y(t)$ respectively, then the system can simply be described by the following linear difference equation:

$$y(t) + a_1 y(t-1) + \cdots + a_n y(t-n) = b_1 u(t-1) + \cdots + b_m u(t-m) \quad (1)$$

The system parameters vector $\theta = \begin{bmatrix} a_1 \cdots a_n \ b_1 \cdots b_m \end{bmatrix}^T$ (2)

and measured input-output data

$$\varphi(t) = \begin{bmatrix} -y(t-1) \cdots -y(t-n) \ u(t-1) \cdots u(t-m) \end{bmatrix}^T \quad (3)$$

Now the estimated system parameters over a time interval ($1 \le t \le N$) can be represented as

$$\hat{\theta} = \min_\theta \left[ \frac{1}{N} \sum_{t=1}^{N} \left( y(t) - \varphi^T(t) \cdot \theta \right)^2 \right] \quad (4)$$

$$\Rightarrow \quad \hat{\theta} = \left[ \sum_{t=1}^{N} \varphi(t) \varphi^T(t) \right]^{-1} \sum_{t=1}^{N} \varphi(t) y(t) \quad (5)$$

Since the sum of the squared residuals or errors (4) is minimized, the method is known as the least square algorithm for system identification [24]. Also with the known value of input and output data at each sampling instant i.e. $\varphi(t)$ vector, using relation (5) the coefficients of the discrete transfer function of the model i.e. $\hat{\theta}$ can be calculated.

### B. Variants of LSE and Choice of a Suitable Estimator

The minimization of the identification error depends largely on the structure of the estimator [22]-[23]. The choice of a suitable structure for the noise model as well as the system model itself plays a very important role in minimization of the prediction error. This sub-section describes few variants of basic LSE and their roles in system identification and choice of a proper estimator.

A generalized linear model structure has the following form

$$y(t) = G(q^{-1}, \theta) u(t) + H(q^{-1}, \theta) e(t) \quad (6)$$

where, $e(t)$ is zero-mean white noise. In (6), $G(q^{-1}, \theta)$ and $H(q^{-1}, \theta)$ are transfer functions of the deterministic and stochastic part of the system respectively. Here, $q^{-1}$ denotes the backward shift operator. Now, (6) can be re-written as

$$A(q^{-1}) y(t) = \frac{B(q^{-1})}{F(q^{-1})} u(t) + \frac{C(q^{-1})}{D(q^{-1})} e(t) \quad (7)$$

where, $\{B, F, C \ \& \ D\}$ are polynomials in $q^{-1}$ and represent the numerator and denominator of the system and noise model respectively. Here, $\{A\}$ represents the polynomial containing the common set of poles for the system and noise model. In [22]-[23], few variants of the generalized LSE can be found depending on the nature of application. It has been seen that the generalized structure (7) can be further customized by considering only fewer number of elements among $\{B, F, C, D \ \& \ A\}$ at once, while choosing different estimators as described below.

*(a) AutoRegressive Exogenous (ARX) Estimator:*

The basic structure of an ARX estimator is defined by:
$$A(q^{-1})y(t) = B(q^{-1})u(t) + e(t) \quad (8)$$
This structure does not allow modeling of the noise and the system dynamics independently. A major disadvantage of this structure is that the deterministic system dynamics and the stochastic noise dynamics both are estimated with the same set of poles which may be unrealistic for many practical cases.

*(b) AutoRegressive Moving Average Exogenous (ARMAX) Estimator:*

Basic structure of an ARMAX estimator is given by:
$$A(q^{-1})y(t) = B(q^{-1})u(t) + C(q^{-1})e(t) \quad (9)$$
The ARMAX structure gives better flexibility over ARX structure to model the measurement noise along with the system. ARMAX structure estimates with different set of zeros but common set of poles for the system and the noise models. This structure is especially suitable when the stochastic dynamics are dominating in nature and the noise enters early into the process e.g. load disturbances.

*(c) Box-Jenkins (BJ) Estimator:*

Basic structure of the BJ estimator is given by the relation:
$$y(t) = \frac{B(q^{-1})}{F(q^{-1})}u(t) + \frac{C(q^{-1})}{D(q^{-1})}e(t) \quad (10)$$
BJ structure allows estimation using different set of poles and zeros for the system and noise model. This model structure is especially suitable when disturbances enter into the model at later stage e.g. measurement noise.

*(d) Output-Error (OE) Estimator:*

An OE estimator has the following structure:
$$y(t) = \frac{B(q^{-1})}{F(q^{-1})}u(t) + e(t) \quad (11)$$

The OE structure estimates poles and zeros for the system model only. It does not estimate the noise model. This structure is suitable when modeling of the system dynamics is the only concern and not the noise-model or the measurement noise is negligible.

### C. Identification of the Reactor Model and Validation

This sub-section presents system identification of a PHWR along with its regulating system using few variants of LSE. For identification, the reactor is visualized as a system with control rod position (fraction of total drop) as input and the global power (in percentage of maximum power produced) as output. The identification is based on data obtained from operating Indian PHWRs provided by the Nuclear Power Corporation of India Ltd. (NPCIL) as also studied in Saha *et al.* [8]. The data at different step-back levels is provided for 14 seconds with 0.1 second of sampling time. Graphical representation of the data is shown in Fig. 2 for 30% and 50% rod drop cases with different initial powers i.e. 100%, 90%, 80% and 70%.

The identification is carried out using System Identification Toolbox [25] of MATLAB and higher order discrete time transfer function models have been estimated from the input output data. The statistical measure of the quality of the identified model can be judged using Akaike's Information Criterion (AIC) [26] and smaller AIC value indicates a better model. Different variants of basic LSE i.e. ARX, ARMAX, BJ and OE are used in the study to find out the most suitable set of models from the test data, having minimum prediction error and the AIC values as compared in Table 1.

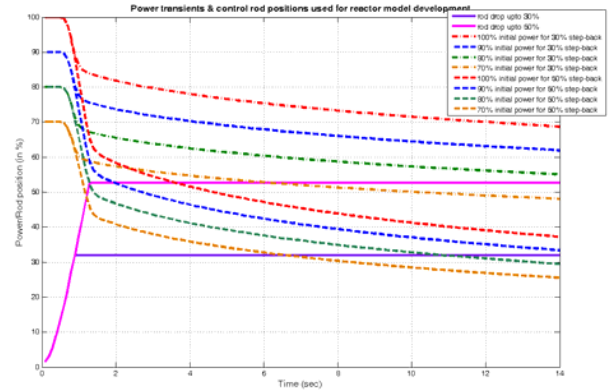

Fig. 2. Power transients and control rod drop data used for reactor modeling.

TABLE 1
CHOICE OF APPROPRIATE ESTIMATOR FOR SYSTEM IDENTIFICATION BASED ON MINIMUM ERROR INDICES (AIC VALUES)

| Identification data for different cases | | Minimum AIC values for different estimators | | | | Preferred structure |
|---|---|---|---|---|---|---|
| Rod drop upto | Initial Power (%) | ARX | ARMAX | BJ | OE | |
| 30% of full length | 100 | -4.3807 | -6.5108 | -6.9891 | -4.3589 | BJ |
| | 90 | -4.5998 | -6.8085 | -7.1948 | -5.6093 | BJ |
| | 80 | -4.7999 | -7.0945 | -7.3565 | -5.2667 | BJ |
| | 70 | -5.0440 | -7.0402 | -7.5382 | -5.0716 | BJ |
| 50% of full length | 100 | -4.0712 | -5.5403 | -6.6792 | -5.4576 | BJ |
| | 90 | -4.3769 | -6.2912 | -7.1863 | -6.1051 | BJ |
| | 80 | -4.6331 | -6.3967 | -7.4852 | -5.2983 | BJ |
| | 70 | -4.7314 | -6.6677 | -7.5635 | -6.1752 | BJ |

It is evident from Table 1 that the Box-Jenkins (BJ) structure is capable of minimizing the prediction error of the reactor model much efficiently over the other structures. The identified discrete time transfer functions using BJ estimator are then converted to continuous time models with the known value of sampling time (0.1 sec). MATLAB's Control System Toolbox [27] function *d2c()* with a customization of discrete to continuous time model conversion with Tustin or bilinear transform has been used. Tustin method enables discrete to continuous time model conversion even with estimated poles, lying very close to the origin of complex z-plane which can not be done with simple Zero-Order Hold (ZOH) type operator. Also, the minimal realization or *minreal()* function [27] has been used which considers any possible pole zero cancellation, thus reducing the order of the identified system and hence model complexity for very close pole-zero estimation in the complex z-plane. The continuous time transfer function models of the nonlinear reactor around different operating points are represented by (12)-(19).

The identified transfer functions for rod drop upto 30%, have been identified as:

$$G_{100}^{30} = \frac{-1.184s^3 - 23.68s^2 + 473.6s + 9472}{s^3 + 11.33s^2 + 55.15s + 48.31} \quad (12)$$

$$G_{90}^{30} = \frac{-1.059s^3 - 21.18s^2 + 423.6s + 8471}{s^3 + 11.35s^2 + 55.17s + 48.02} \quad (13)$$

$$G_{80}^{30} = \frac{-0.9666s^3 - 19.33s^2 + 386.7s + 7733}{s^3 + 11.52s^2 + 55.93s + 49.15} \quad (14)$$

$$G_{70}^{30} = \frac{-0.8387s^3 - 16.77s^2 + 335.5s + 6710}{s^3 + 11.43s^2 + 55.87s + 49.11} \quad (15)$$

Here, the subscripts of transfer function model $G$ denote the initial reactor power and the superscripts denote the rod drop level. In a similar manner, the transfer functions for rod drops upto 50% have also been identified as:

$$G_{100}^{50} = \frac{\begin{array}{c}-0.1376s^5 - 8.258s^4 - 110.1s^3 \\ +2202s^2 + 66070s + 440400\end{array}}{s^5 + 11.24s^4 + 236.8s^3 + 1584s^2 + 6729s + 8031} \quad (16)$$

$$G_{90}^{50} = \frac{-0.148s^4 - 5.919s^3 + 2368s + 23680}{s^4 + 14.4s^3 + 108.3s^2 + 396.4s + 442} \quad (17)$$

$$G_{80}^{50} = \frac{-0.1934s^3 - 3.869s^2 + 77.38s + 1548}{s^3 + 7.994s^2 + 33.19s + 34.01} \quad (18)$$

$$G_{70}^{50} = \frac{-0.1894s^3 - 3.788s^2 + 75.75s + 1515}{s^3 + 8.297s^2 + 36.08s + 38.72} \quad (19)$$

In (12)-(19), the identified models of the PHWR are found to be different with each other since reactor dynamics is inherently nonlinear in nature and its linearized model will differ depending on the operating point [8], [9]-[10]. Indeed a PHWR has a pole in the open loop mode as in [7], whereas the reactor has been modeled in closed loop along with its power regulator as shown in Fig. 1, which is a stable system as reported in equation (12)-(19). Such modeling enables to design a single iso-damped FOPID controller that can be incorporated with minimum change in the existing RRS loop.

It is also worth mentioning that for identification of nonlinear processes as linearized transfer functions around different operating points, there is no true model that can be derived analytically from the governing physical laws. In most cases, linearized modeling of systems is done for a chosen input excitation. In a nonlinear system like the test PHWR, the maximum order of the linearized model can not be determined from the governing laws (i.e. point kinetics). In such cases, the order selection is done iteratively by gradually increasing the maximum order of the model and hence system complexity while also comparing the prediction error (AIC values) of the estimated model with that one in previous iteration until it mimics the original dynamic behavior of the plant around a specific operating point. In the present study, the main objective of reactor transfer function modeling is to represent the nonlinear behavior of the reactor as a set of linear models having minimum prediction error for a specific input excitation (truncated ramp in this case in Fig. 2) while also keeping the model order as low as possible.

## III. FRACTIONAL ORDER MODELING OF THE PHWR

### A. Need of a FO Model Reduction Template for Robust Tuning of $PI^\lambda D^\mu$ Controllers

Since, FOPID tuning needs a reduced order template with sufficient accuracy, a FO model reduction technique, proposed in [13] is applied for the reduction of the identified models of the PHWR around different operating points, given by (12)-(19). In the present study, the identification is carried out with both 30% and 50% rod drop data using four variants of LSE among which Box-Jenkins structure has been found to have better accuracy over the other variants (Table 1). Thus it is logical that the accuracy so obtained with a proper choice of estimator in system identification should be retained in the reduced parameter modeling also. In order to do so, model reduction in *flexible order* structure has been attempted next. The new reduced parameter templates are introduced below:

*(a) One Non-integer Order Plus Time Delay (NIOPTD-I):*

$$P_1(s) = \frac{K}{Ts^\alpha + 1} e^{-Ls} \quad (20)$$

*(b) Two Non-integer Orders Plus Time Delay (NIOPTD-II):*

$$P_2(s) = \frac{K}{s^\alpha + 2\zeta\omega_n s^\beta + \omega_n^2} e^{-Ls} \quad (21)$$

In equation (20) and (21), the system parameters are defined by the analogous pseudo-parameters of the conventional FOPTD and SOPTD structures i.e. time constant ($T$), damping ratio ($\zeta$), natural frequency ($\omega_n$), propagation delay ($L$) and dc-gain ($K$) with flexible orders $\{\alpha, \beta\}$. These templates take the classical FOPTD and SOPTD structures for $\alpha = 1, \beta = 1$, respectively.

Saha *et al.* [8] reduced the identified models in FOPTD and SOPTD template by minimizing the difference between $H_2$-norm of the identified and reduced order model as proposed by Xue & Chen [28]. In the present study, model reduction in NIOPTD-I and NIOPTD-II templates are attempted with the minimization of the $H_2$-norm of the identified and the fractional order NIOPTD models (20)-(21). The $H_2$ norm of a system reflects how much it amplifies or attenuates its inputs over all the frequencies. In other words, it represents the energy of the output signal of a system, subjected to an impulse excitation. Mathematically, $H_2$ norm of a system $P(s)$ can be evaluated by the following relation

$$\|P(s)\|_2 = \sqrt{\frac{1}{2\pi} \int_{-\infty}^{\infty} trace\left[P(j\omega)\overline{P(j\omega)^T}\right] d\omega} \quad (22)$$

Here, $J = \|P(s) - \tilde{P}(s)\|_2 \quad (23)$

The objective function ($J$) in the model reduction technique has been minimized with unconstrained Nelder-Mead Simplex algorithm implemented in MATLAB's Optimization Toolbox [29] function *fminsearch()* with perturbed initial guesses. The algorithm searches for an optimal set of model parameter $\{K, T, L, \alpha\}$ for (20) and $\{K, \zeta, \omega_n, L, \alpha, \beta\}$ for (21) while minimizing the deviation in 2-norm of those two systems. The model reduction technique proposed in [28], often fails as any unstable mode, encountered within the search space will immediately stop the optimization process and also due to the $H_2$ norm of an unstable system being infinite. As a solution, the search is restricted with the MATLAB functions $(isfinite(P)==1)$, $(isa(P,"tf")==1)$ and $(isstable(P)==1)$ to avoid infinite, dc-modes with pole-zero cancellation and most significantly unstable modes in the model reduction process. Also, a large penalty function has been included within the objective function (23) which strongly discourages parameter search with unstable modes. Also, in each iteration of the optimization process, the guess values of the FO elements are continuously rationalized with a fourth order Oustaloup's approximation within the frequency range of $\omega \in [10^{-4}, 10^4]$ rad/s. The inherent delays within the identified non-minimum phase reactor models (12)-(19) are also extracted by the proposed technique with an equivalent 3$^{rd}$ order Pade approximation.

### B. FO Model Reduction of the Identified System

The model reduction technique presented in the previous sub-section has now been applied to the identified reactor models (12)-(19). The accuracy of the reactor models as the minimized value of the objective function, are compared for the four classes of reduced order templates viz. FOPTD, SOPTD, NIOPTD-I and NIOPTD-II in Table 2. It is clear from Table 2 that NIOPTD-II template is capable of capturing the linearized higher order dynamics of the nonlinear reactor much efficiently over the other three templates.

TABLE 2
CHOICE OF APPROPRIATE REDUCED ORDER TEMPLATE BASED ON MINIMUM VALUE OF THE OBJECTIVE FUNCTION

| Identification data for different cases | | DC Gain | Minima of the modeling error (normalized by DC gain) for different reduced order templates | | | |
|---|---|---|---|---|---|---|
| Rod drop level | Initial Power (%) | | FOPTD | SOPTD | NIOPTD-I | NIOPTD-II |
| 30% of full length | 100 | 196.063 | 3.5402 | 0.2542 | 2.6049 | 0.1220 |
| | 90 | 176.406 | 3.1654 | 0.2290 | 2.3416 | 0.1105 |
| | 80 | 157.339 | 2.8988 | 0.2008 | 2.1348 | 0.1782 |
| | 70 | 136.641 | 2.5229 | 0.1910 | 1.8904 | 0.0935 |
| 50% of full length | 100 | 58.246 | 3.2987 | 0.4507 | 2.3918 | 0.3979 |
| | 90 | 53.558 | 2.7642 | 0.4206 | 1.9714 | 0.1388 |
| | 80 | 45.509 | 2.0806 | 0.2694 | 1.5163 | 0.0872 |
| | 70 | 39.125 | 1.8682 | 0.2556 | 1.3874 | 0.0696 |

The reduced parameter reactor models (in generalized fractional order templates) are reported in Table 3, corresponding to the lowest modeling errors in Table 2. The large variation in the dc gain of the models (around different operating points) is also evident from Table 2-3. It is also observed that for all the NIOPTD-I class of models $T$ and $L$ are almost constant but the gain ($K$) is rapidly varying with shift in operating point. Similar nature can be observed for the NIOPTD-II models also, where it is found the variation in $\zeta$, $\omega_n$ and $L$ to be negligibly small compared to the variation in $K$ with operating point shifting.

TABLE 3
REDUCED FRACTIONAL ORDER MODELS OF THE REACTOR AROUND DIFFERENT OPERATING POINTS

| Identified Models | Class of Models | Reduced Fractional Order Models |
|---|---|---|
| $P_{100}^{30}$ | NIOPTD-I | $\dfrac{195.0736}{1.0006s^{1.057}+1}e^{-0.0934s}$ |
| | NIOPTD-II | $\dfrac{1522.8947}{s^{2.0971}+8.1944s^{1.0036}+7.7684}e^{-2.0043\times10^{-12}s}$ |
| $P_{90}^{30}$ | NIOPTD-I | $\dfrac{175.5107}{1.0084s^{1.0559}+1}e^{-0.0937s}$ |
| | NIOPTD-II | $\dfrac{1359.2345}{s^{2.0972}+8.1906s^{1.0036}+7.7075}e^{-1.5968\times10^{-9}s}$ |
| $P_{80}^{30}$ | NIOPTD-I | $\dfrac{156.5649}{0.99866s^{1.0566}+1}e^{-0.0928s}$ |
| | NIOPTD-II | $\dfrac{1027.3027}{s^{2.0163}+6.7859s^{0.99388}+6.5268}e^{-2.5346\times10^{-5}}$ |
| $P_{70}^{30}$ | NIOPTD-I | $\dfrac{136.0175}{0.99724s^{1.0546}+1}e^{-0.0932s}$ |
| | NIOPTD-II | $\dfrac{1074.396}{s^{2.0961}+8.2663s^{1.0037}+7.8641}e^{-3.1431\times10^{-10}s}$ |
| $P_{100}^{50}$ | NIOPTD-I | $\dfrac{57.8883}{0.72536s^{1.1112}+1}e^{-0.0746s}$ |
| | NIOPTD-II | $\dfrac{529.1365}{s^{2.1002}+7.1111s^{1.0002}+9.0873}e^{-1.6049\times10^{-5}s}$ |
| $P_{90}^{50}$ | NIOPTD-I | $\dfrac{53.3154}{0.64365s^{1.0958}+1}e^{-0.1475s}$ |
| | NIOPTD-II | $\dfrac{604.2541}{s^{2.2986}+8.871s^{1.0321}+11.2993}e^{-6.5\times10^{-6}s}$ |
| $P_{80}^{50}$ | NIOPTD-I | $\dfrac{45.3026}{0.79668s^{1.0849}+1}e^{-0.1379s}$ |
| | NIOPTD-II | $\dfrac{337.846}{s^{2.2038}+6.7453s^{1.0132}+7.4275}e^{-1.8479\times10^{-7}s}$ |
| $P_{70}^{50}$ | NIOPTD-I | $\dfrac{38.957}{0.7557s^{1.0836}+1}e^{-0.1291s}$ |
| | NIOPTD-II | $\dfrac{325.2142}{s^{2.1969}+7.1459s^{1.0113}+8.3167}e^{-2.343\times10^{-7}s}$ |

Here, all the reduced parameter models extract the inherent time delays of the identified non-minimum phase reactor models (12)-(19) along with the dominant fractional orders [17] which should not be restricted to take integer orders only like classical FOPTD and SOPTD modeling. The reduced parameter modeling of the reactor is done with the motive that the nonlinear dynamics of the plant can easily be represented by a set of linear varying gain systems in an accurate NIOPTD template. Next, simulated time responses of the NIOPTD-II models under step-back are presented in Fig. 3 for model validation.

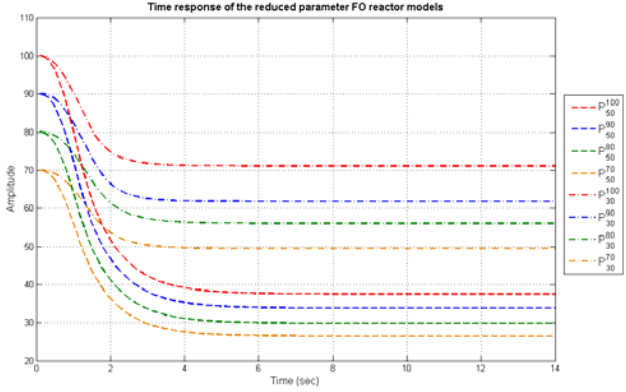

Fig. 3. Reduced model validation under different step-back condition.

## IV. DESIGN OF ROBUST FOPID AND PID CONTROLLERS FOR ACTIVE STEP-BACK IN THE NUCLEAR REACTOR

The reactor model (in the most accurate NIOPTD-II template in Table 3) is first tuned with the FOPID controller having the following structure:

$$C(s) = K_p + \frac{K_i}{s^\lambda} + K_d s^\mu \quad (24)$$

Here, $\{K_p, K_i, K_d\}$ are the proportional, integral and derivative gains and $\{\lambda, \mu\}$ are the orders of differ-integrals of the $PI^\lambda D^\mu$ controller. Clearly, for $\lambda=1, \mu=1$ controller (24) takes the classical PID form. The performance of the $PI^\lambda D^\mu$ controller for the control of reduced order reactor models is expected to be better than the classical PID controller due to its higher degrees of freedom for tuning. But, a systematic tuning strategy for FOPID controller is required that would be most suitable to handle the wide variation in reactor model with initial power and rod drop level. Hence, available FOPID tuning techniques [30] are further enhanced as in [13] which require lesser computational load, achieve high robustness while meeting other specifications like noise rejection, load disturbance rejection levels etc.

The concept of frequency domain design of FOPID controllers was first proposed by Monje *et al.* [30]. If $P(s)$ be the transfer function of the process, then the objective is to find out a controller $C(s)$, so that the open loop system $G(s) = C(s)P(s)$ meets the following design specifications:

*(a) Phase margin specification:*
$$Arg[G(j\omega_{gc})] = Arg[C(j\omega_{gc})P(j\omega_{gc})]$$
$$= -\pi + \phi_m \quad (25)$$

*(b) Gain crossover frequency specification:*
$$|G(j\omega_{gc})| = |C(j\omega_{gc})P(j\omega_{gc})| = 1 \quad (26)$$

*(c) Robustness against system's gain variation (Iso-damping):*
$$\left(\frac{d}{d\omega}(Arg[G(j\omega_{gc})])\right)_{\omega=\omega_{gc}} = 0 \quad (27)$$

*(d) Complementary Sensitivity specification:*

$$|T(j\omega)| = \left|\frac{C(j\omega)P(j\omega)}{1+C(j\omega)P(j\omega)}\right|_{dB} \leq A\ dB \quad \forall \omega \geq \omega_t\ rad/s$$
$$\Rightarrow |T(j\omega_t)| = A\ dB \quad (28)$$

where, $A$ is the specified magnitude of the complementary sensitivity function or noise attenuation for frequencies $\omega \geq \omega_t\ rad/s$.

*(e) Sensitivity Specification:*
$$|S(j\omega)| = \left|\frac{1}{1+C(j\omega)P(j\omega)}\right|_{dB} \leq B\ dB \quad \forall \omega \leq \omega_s\ rad/s$$
$$\Rightarrow |S(j\omega_s)| = B\ dB \quad (29)$$

where, $B$ is the specified magnitude of the sensitivity function or load disturbance rejection for frequencies $\omega \leq \omega_s\ rad/s$.

So, the five parameters of the $PI^\lambda D^\mu$ controller (24) can now be tuned using the five specifications (25)-(29) to get a unique solution. Here, Powell's Trust-Region-Dogleg algorithm, implemented in MATLAB's Optimization Toolbox [29] function *fsolve()* is used to find out the values of the controller parameters $\{K_p, K_i, K_d, \lambda, \mu\}$.

It may be mentioned that PID controllers can also be designed for FO plants as reported in [31]. It is therefore worthwhile to compare the performance of the plant represented as a NIOPTD system with a PID and $PI^\lambda D^\mu$ controller. Hence, a PID controller is tuned for comparison with the FOPID controller using the same optimization technique using only three specifications (25)-(27), since it has only three tuning parameters. The tuning is attempted with the highest gain model, 30% drop at 100% power i.e. $P_{100}^{30}$ of NIOPTD-II structure (Table 3) for a PID and FOPID controller. The $PI^\lambda D^\mu$ and PID controllers tuned with an over-damped phase margin specification and $\omega_{gc} = 1\ rad/\sec$ are:

$$C^{FOPID}(s) = 0.0006 + \frac{0.0052}{s^{1.0137}} + 0.0049 s^{0.1067} \quad (30)$$

$$C^{PID}(s) = 0.0052 + \frac{0.0051}{s} + 0.00007s \quad (31)$$

The corresponding flat phase curve of the open loop system comprising of the $P_{100}^{30}$ reactor model of NIOPTD-II template and controllers (30)-(31) are shown in Fig. 4. The rationale behind attempting iso-damped tuning with the reactor models in Table 3, are discussed next. Saha *et al.* [8] tuned a FO phase shaper augmented with a PID controller for the slowest FOPTD/SOPTD reactor model as the FOPTD and SOPTD systems representing the PHWR under a step-back at different power levels were systems with varying dc-gain only with the gain increasing with the power level at which the step-back is initiated. The present work tunes more accurate NIOPTD-II models having the maximum gain which represents the system identified for 30% drop at 100% power. The rationale behind this is explained as follows, starting with Table 3.

In Table 3, for the reduced parameter models corresponding to equations (12)-(19), the time-delays are much accurately extracted in the flexible NIOPTD-II templates which were same for the SOPTD based modeling in [8]. This is the cause of deviation in the phase curves of the reactor models above a certain frequency as shown in Fig. 5 although the phase curve is almost same at lower frequencies. It is seen further that the phase margin reduces with increased initial power for the same bulk change in reactivity of the system representing the plant and also with the decrease in the quantum of bulk power change, for a given initial power level. Thus, the reduced system corresponding to 30% change at 100% power has the minimum phase margin and the maximum dc gain and may be viewed as the fastest system.

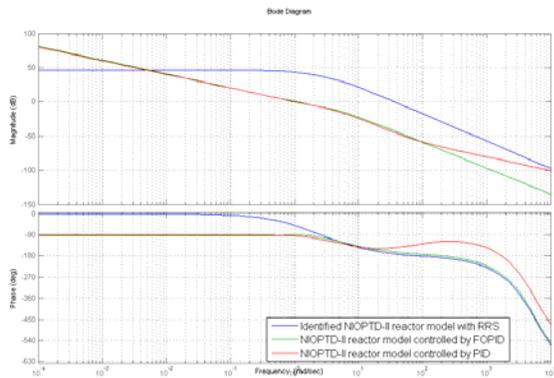

Fig. 4. Phase flattening of the NIOPTD-II reactor model with FOPID and PID.

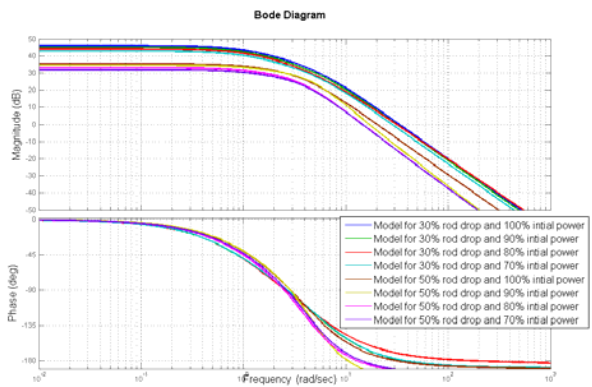

Fig. 5. Bode diagram of the NIOPTD-II models of the PHWR.

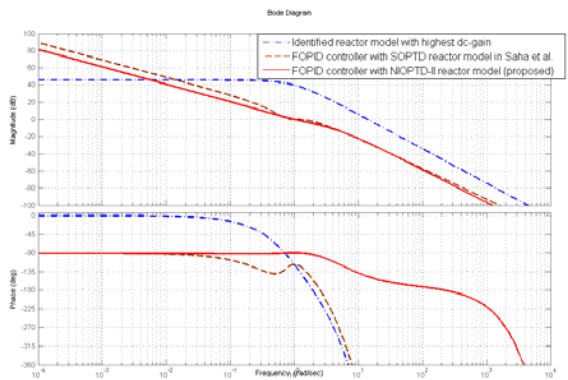

Fig. 6. comparison of flatness in phase curve with SOPTD [8] and the proposed NIOPTD-II modeling.

An examination of Fig. 5 shows further that the phase curves for all the reduced order systems are almost overlapping and have the same slope up to a certain frequency of about 2 rad/sec and then proceed towards their asymptotic values along almost parallel trajectories. Thus, if these phase curves are flattened using a FOPID controller around a frequency $\omega_{gc}$ in the overlapping region and the gain-crossover frequency of the open-loop system with the controller is fixed at this point, then the closed loop system will exhibit iso-damped response for gain variation as the phase-margin remains constant. Since the fastest system has the minimum phase margin, a controller which meets the phase margin specifications for the fastest system is sure to produce a better or equal phase margin for the slower systems. The relationship between gain of a system, its phase margin and slope of its phase curve can be expressed using Bode's Integral and has been explained in [8] for robust controller design. Also, in Table 3, the identified orders of the NIOPTD models are close to 1 and 2 in some cases suggesting a simpler SOPTD modeling. The adverse effect of loosing modeling accuracy on the controller design is highlighted in Fig. 6. It shows that the robust frequency domain tuning of FOPID controllers make a wide flat phase curve in the lower frequencies rather than the higher frequencies and the width of the flatness is higher for the NIOPTD-II based design rather than the SOPTD based design as in Saha *et al.* [8]. Indeed the modeling inaccuracies due to reduction in SOPTD template as reported in [8] minimizes the achievable phase-margin as evident from Fig. 6, although the corresponding phase curves are locally flat around $\omega_{gc}$. Further, the parametric robustness achieved with the present tuning methodology is much more compared to that achieved by Saha *et al.* [8] as seen from the width of the flat phase region around $\omega_{gc}$. This is particularly important for the present problem of controlling a reactor where the system's dc-gain changes with power level. Since, the SOPTD templates estimates the system delay to be in higher magnitude than they actually are [13] which makes the phase curve to droop much earlier causing lesser phase margin. Modeling in NIOPTD-II template reduces the modeling error to a significant extent causing preservation of the desired phase margin, with flat phase around a desired $\omega_{gc}$ that is an improvement over [8]. As mentioned in [8], the increased parametric robustness is achieved at the cost of reduction in phase margin. The methodology uses a constrained optimization which maximizes the region of flatness with the minimum phase margin specified as a constraint as elaborated in [8]. The physical implication is that the system without the phase shaper must be tuned to a high value of closed loop damping so that the phase margin with the phase shaper in loop is appreciable. This drawback is eliminated in the present approach.

This justifies tuning of the fastest NIOPTD-II reactor model ($P_{100}^{30}$) for tuning with a robust FOPID controller. The concept can be viewed as if the gain of the robust FOPID, tuned with the fastest plant, is reduced to handle the reactor dynamics at other operating points (for slower plants) and can also be increased at per wish to get a faster control action, keeping the power undershoot same. Fig. 7 shows the iso-damped response of all the reactor models around different operating points. As, the fastest plant has been controlled by a robust frequency domain tuning technique, the other reactor models, having much lower dc-gains and slower transient responses get automatically handled with the $PI^\lambda D^\mu$ and integer order PID controllers. In Fig. 8 the loop gain has been increased to get faster control action for rod drop within the reactor with an intention of keeping the overshoot at same level. A simple PID has been shown to produce power undershoot at 350% increase in the loop gain whereas the FOPID maintains a dead-beat response. From Fig 7-8, it is evident that with a robust FOPID controller, the dc gain of the fastest reactor model ($P_{100}^{30}$) can be decreased even upto 500% to handle the change in linearized models and can be increased also upto 350% to achieve faster control action, due to its higher capability of enforcing iso-damping property than a simple integer order PID controller.

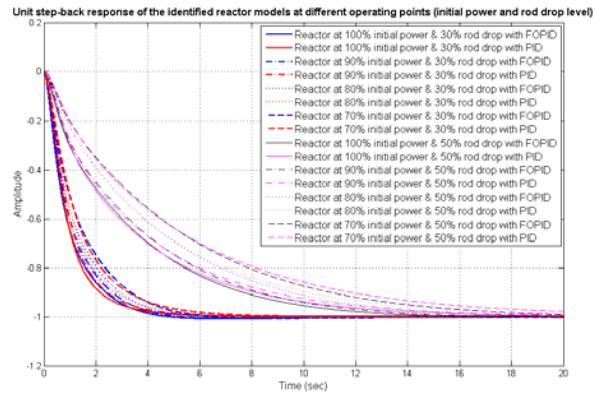

Fig. 7. Unit step-back response of the identified NIOPTD-II reactor models at different operating points (i.e. initial power and rod drop level).

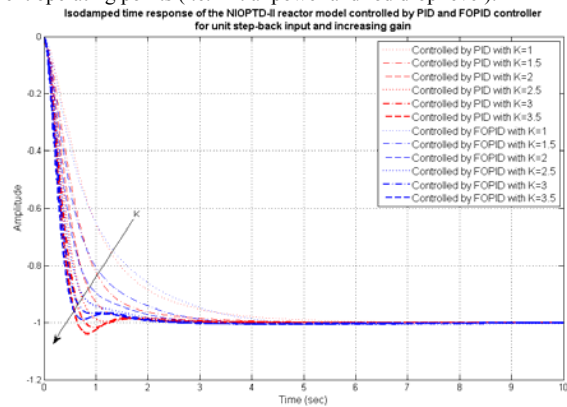

Fig. 8. Iso-damped time response of the NIOPTD-II reactor models with FOPID & PID controllers for unit step-back input with 350% increase in gain.

## V. SIMULATIONS AND RESULTS OF THE ACTIVE STEP-BACK FOR RAPID POWER REDUCTION IN THE PHWR

The power transients due to the active rod drop are now simulated with the FOPID and PID controller, obtained in the previous section. The reactor, initially operating at 100%, 90%, 80% and 70% of full power is tested with the 30% and 50% step back, with the proposed robust tuning of FOPID and PID controllers. The proposed FOPID controllers shows dead-beat tracking performance of the desired power level (Fig. 9-10) with high degree of robustness against change in linearized reactor models (Fig. 7-8) due to nonlinearity and deliberate gain variation for faster control action. Fig. 9-10 clearly shows that the robust FOPID performs much better over the present RRS in handling sluggish step-back response for truncated ramp input (Fig. 2).

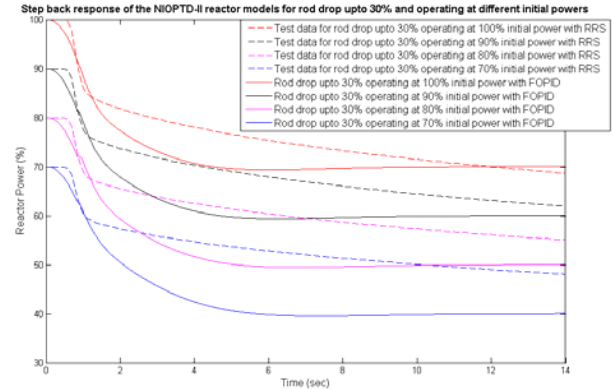

Fig. 9. 30% step back response of the reactor models at various initial powers.

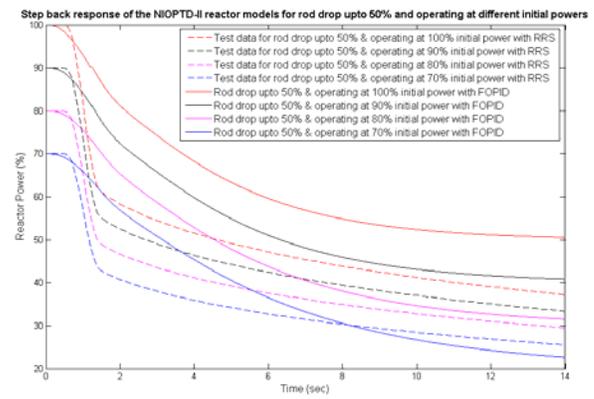

Fig. 10. 50% step back response of the reactor models at various initial powers.

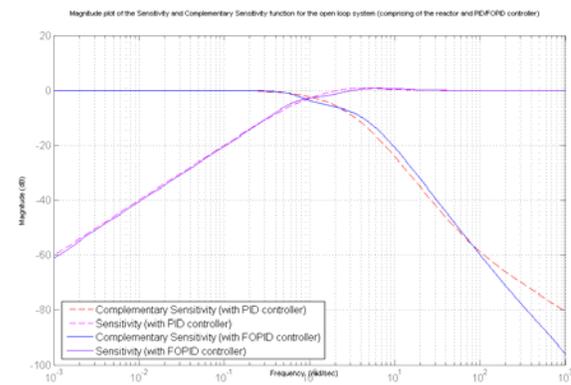

Fig. 11. Sensitivity and complementary sensitivity of the open loop system.

It is also evident that the present RRS with a proportional only controller shows steady-state error, due to the low proportional gain at low powers. As the gain increases, the offset reduces gradually. The presence of the fractional order integrator of the proposed FOPID controller (30) forces the steady-state offsets of power transient to zero. The other specifications of FOPID design like maximum magnitude of sensitivity and complementary sensitivity function, are also satisfied which is evident from Fig. 11. Clearly, $|S(j\omega)|$ at $\omega_s = 10^{-2} \, rad/s$ and $|T(j\omega)|$ at $\omega_t = 10^2 \, rad/s$ are lesser for a $PI^\lambda D^\mu$ controller than a simple PID controller. This shows the advantage of a PI$^\lambda$D$^\mu$ controller for load disturbance suppression and high frequency measurement noise rejection compared to a PID controller.

From the above simulation studies, it can be concluded that several limitations of the passive step-back mechanism of the operating Indian PHWRs like steady-state off-set, change in linearized models due to high nonlinearity, sluggish transient response, large power undershoot for sudden negative reactivity insertion can be efficiently handled with the technique proposed in this paper. Thus, a FOPID controller due to having better flexibility or higher number of parameters for tuning is most suitable for operating in conjunction with the present RRS with the scheme shown in Fig. 1, rather than a PID controller. Hardware implementation issues of such FOPID controllers are discussed in [13], [17] which includes fractance, analog electronic circuit based realization; FPGA based digital realization and electrochemical realization by lossy capacitors etc.

## VI. CONCLUSION

In this paper, the nonlinear process dynamics of an operating PHWR has been modeled as several linearized transfer function models from practical test-data with standard variants of LSE, around various operating points. The identified models are reduced in fractional order templates namely NIOPTD-I and NIOPTD-II which capture the identified dynamics of the reactor much better than the classical FOPTD and SOPTD templates and are most suitable for tuning with a PI$^\lambda$D$^\mu$ controller. A robust frequency domain tuning technique is adopted, based on simultaneous nonlinear equation solving for tuning of the FOPID controller. The methodology, put forward in this paper, outperforms the RRS in its present form for the Indian PHWRs. This technology can be used for enhancing the parametric robustness of the existing RRS control loops with variation in system's gain. The iso-damped nature of the closed loop system allows much faster dead-beat power tracking under a step-back condition. This concept can be extended to identification of the test PHWR as continuous order models in future works.